# A Hidden Quantum Paraelectric Phase in $SrTiO_3$ Induced by Terahertz Field

Wei Li[1,‡], Hanbyul Kim[2,‡], Xinbo Wang[3,‡], Jianlin Luo[3], Simone Latini[4,5], Dongbin Shin[2,4,†], Jun-Ming Liu[6], Jing-Feng Li[1], Angel Rubio[4,7], Ce-Wen Nan[1], Qian Li[1,*]

[1]*State Key Laboratory of New Ceramics and Fine Processing, School of Materials Science and Engineering, Tsinghua University, Beijing 100084, China.*
[2]*Department of Physics and Photon Science, Gwangju Institute of Science and Technology (GIST), Gwangju 61005, South Korea.*
[3]*Beijing National Laboratory for Condensed Matter Physics, Institute of Physics, Chinese Academy of Sciences, Beijing 100190, China.*
[4]*Max Planck Institute for the Structure and Dynamics of Matter and Center for Free-Electron Laser Science, Luruper Chaussee 149, 22761 Hamburg, Germany.*
[5]*Department of Physics, Technical University of Denmark, 2800 Kgs. Lyngby, Denmark.*
[6]*Laboratory of Solid State Microstructures and Innovation Center of Advanced Microstructures, Nanjing University, 210093, Nanjing, China.*
[7]*Center for Computational Quantum Physics (CCQ), The Flatiron Institute, 162 Fifth Avenue, New York, New York 10010, USA.*

Coherent manipulation of lattice vibrations using ultrafast light pulses enables access to non-equilibrium 'hidden' phases with designed functionalities in quantum materials. However, expanding the understanding of nonlinear light-phonon interaction mechanisms remains crucial for developing new strategies. Here, we report re-entrant ultrafast phase transitions in SrTiO3 driven by intense terahertz excitation. As the terahertz field increases, the system transitions from the quantum paraelectric (QPE) ground state to an intermediate ferroelectric phase, and then unexpectedly reverts to a QPE state above ~500 kV/cm. The latter hidden QPE phase exhibits distinct lattice dynamics compared to the initial phases, highlighting activated antiferrodistortive phonon modes. Aided by first-principles dynamical calculations, we identify the mechanism for these complex behaviors as a superposition of multiple coherently excited eigenstates of the polar soft mode. Our results reveal a previously uncharted quantum facet of SrTiO3 and open pathways for harnessing high-order excitations to engineer quantum materials in the ultrafast regime.

*Introduction*—Manipulating crystal structures using ultrafast light pulses has emerged as a powerful approach to engineering the properties of quantum materials, surpassing the limitations of static structural control methods [1–5]. These engineered properties, often arising from nonequilibrium states on picosecond timescales, hold potential for applications in quantum and neuromorphic computing [6], as well as next-generation optoelectronics [7–9]. A prominent research strategy has been termed nonlinear phononics, where resonant pumping of an infrared (IR)-active phonon mode at mid-IR frequencies initiates the excitation of a coupled Raman-active mode [10]. Such coupled dynamics induce transient structural distortions, as dictated by the Raman mode, endowing properties such as high-temperature superconductivity [11,12] and ferrimagnetism [13]. Whereas phonon-phonon interactions have been addressed in these studies, the nonlinearities inherent in individual phonon modes and their dynamical effects are largely underutilized [14–16].

In parallel, terahertz (THz) pulses have been employed to directly excite low-lying soft modes, driving collective ion motions far from equilibrium. As an archetypal perovskite material, $SrTiO_3$ exhibits a rich spectrum of ultrafast THz-induced behaviors [17–21]. Under thermal equilibrium, $SrTiO_3$ undergoes a phase transition from a paraelectric (PE) cubic structure to an antiferrodistortive (AFD) tetragonal structure at a critical temperature, $T_A \sim 105$ K [22]. Further cooling below the Curie temperature, $T_C \sim 36$ K, leads to a quantum paraelectric (QPE) ground state, rather than long-range ferroelectric (FE) order, due to zero-point quantum fluctuations [23,24]. With multiple competing structural instabilities, including lattice strain, AFD distortions, and FE ordering [25,26], the QPE phase provides a sensitive platform for structural control via strain tuning and chemical doping [27–29]. Upon intense THz electric field excitation, a non-equilibrium, 'hidden' FE phase has been observed in QPE $SrTiO_3$, as indicated by second harmonic generation (SHG) probe signals featuring a non-oscillatory, decaying component

*Contact Author: qianli_mse@tsinghua.edu.cn
†Contact Author: dshin@gist.ac.kr
‡These authors contributed equally to this work.

over ~10 picoseconds. However, recent studies on another QPE system, $KTaO_3$, demonstrated that non-oscillatory SHG signals can also originate from THz-induced dipole correlations, casting doubt on their robustness as a fingerprint of the transient FE phase [30,31]. Thus, despite the observation of hidden phases with exotic properties, a comprehensive understanding, particularly of the quantum mechanical nature of these THz-induced phase transitions, remains elusive.

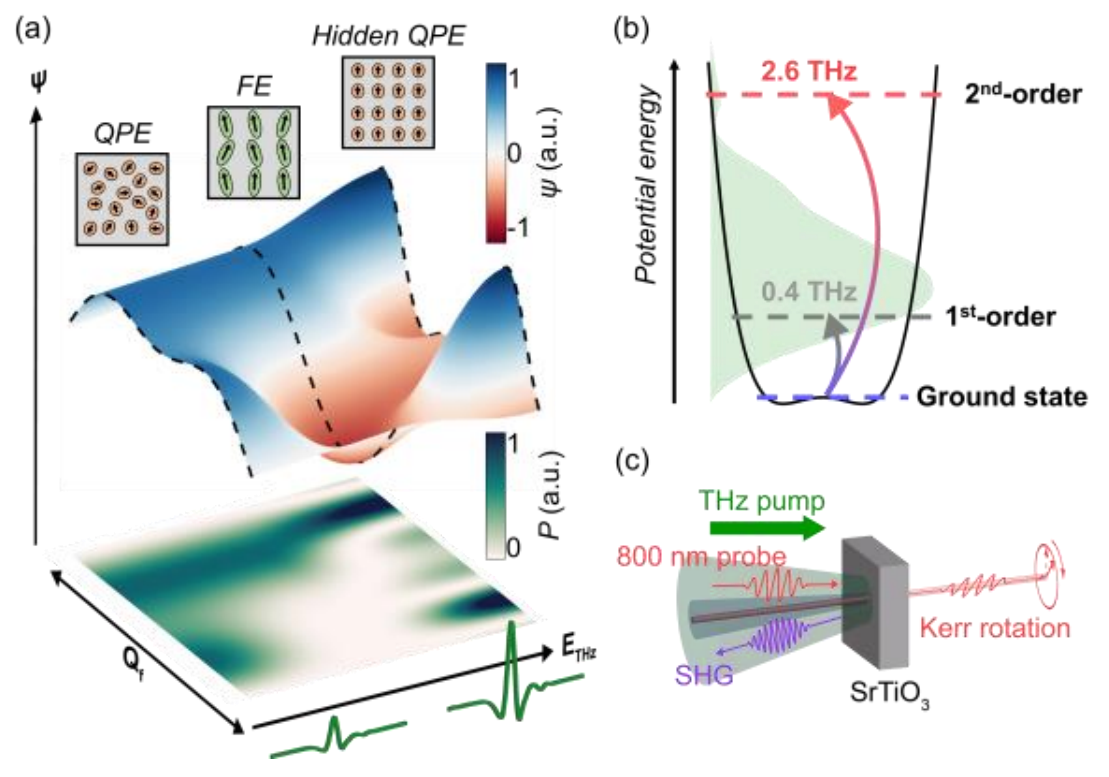

FIG. 1. Emergence of the hidden QPE phase in $SrTiO_3$ under THz excitation. (a) Schematic of the wave functions (*Ψ*) along the soft-mode coordinate ($Q_f$) for different THz-induced states, together with the projected probability (*P*) distribution map. Three typical THz field strengths are marked with dashed lines corresponding to the transitions of the QPE, FE and hQPE phases. Insets illustrate the dipole configurations of these phases. (b) The potential energy function of $SrTiO_3$ with respect to $Q_f$, showing the calculated first-order and second-order eigenmode frequencies of ~0.4 THz and ~2.6 THz, respectively. The experimental THz excitation spectrum is overlaid in a green shadow. (c) Illustration of TFISH and TKE measurements. THz pump pulses were incident normally onto a (001)-cut $SrTiO_3$ single crystal, with their polarization direction set parallel to the [100] crystallographic direction. The polarization direction of collinearly incident 800 nm probe pulses were set at 0° and 45° in TFISH and TKE measurements, respectively, relative to the THz polarization. TFISH was measured in a near-backscattering geometry while TKE in the transmission geometry using a balanced detection scheme.

Here, we explore high-order nonlinear phonon excitation as a strategy to uncover hidden phases in quantum materials. By resonantly exciting its soft mode, we observed a THz-induced QPE-FE transition in $SrTiO_3$, succeeded by a re-entrant hidden QPE (hQPE) phase at elevated THz field strengths. Combining first-principles dynamical simulations, we demonstrate that excitation of the soft mode to its second-order state triggers the hQPE phase transition and the activation of an enhanced AFD mode. Comparative analyses of the phonon evolution behaviors suggest a complex interplay between the soft mode and AFD-type modes across the first- and second-order excited states. Our findings deepen the understanding of phonon nonlinearity and resolve the controversy over THz-induced ferroelectricity in QPE systems.

*High-order excitation mechanism and observations*—Fig. 1A illustrates the schematic of the high-order excitation mechanism in $SrTiO_3$, adapted from the first-principles calculations (FIG. S1 [32]). $SrTiO_3$ features a shallow double-well potential energy landscape along the soft mode coordinate $Q_f$ which represents the out-of-phase motion of Ti-ions relative to oxygen octahedra. In the ground state, the $Q_f$ wave function forms a double-well with a nearly flat top, resulting in a symmetric probability distribution across both wells. This wave function dictates a macroscopically nonpolar structure known for the QPE phase. In the first-order excited state, the wave function becomes anti-symmetric, with opposite signs in the positive and negative $Q_f$ regions. When subjected to a moderate THz electric field, $SrTiO_3$ forms a superposition of the ground state and the first-order excited state, causing the probability density to shift towards one side of $Q_f$. This symmetry breaking is associated with the long-range ordering of electric dipoles, triggering a QPE to FE phase transition. At this stage, the proportion of the FE phase grows with increasing the THz field, in correlation with the phonon mode population or its dipole moment. Beyond a critical field strength, the second-order excited state becomes significant, where the wave function adopts a prominent, symmetric double-peak shape along $Q_f$ (FIG. S1 [32]). At this point, a coherent mixture of the ground, first-order and second-order excited states emerges, with their relative proportions depending on the THz pulse spectrum. As the THz field continues to increase, the increasingly populated second-order excited state restores the symmetry of the $Q_f$ probability density distribution. These soft-mode dynamics nullify the induced dipole moment, resulting in a re-entrant phase transition from the FE phase to a previously unknown hQPE phase.

To confirm the QPE-FE-hQPE phase transition pathway in $SrTiO_3$, we performed ultrafast pump-probe spectroscopy using intense THz pulses generated via a $LiNbO_3$ crystal (FIG. S3 [32]). The THz pulses were found to reach a peak electric field in excess of 1000 kV/cm, covering the spectral range of ~0.1–2.6 THz (Fig. 1B). We employed two optical probes based on THz field-induced SHG (TFISH) and Kerr rotation effects (TKE), as illustrated in Fig. 1C. TFISH is sensitive to the lattice symmetry breaking, while TKE reflects changes in the linear optical susceptibility (at 800 nm here). Together, these experiments provide comprehensive information on the THz-induced dynamics of both IR- and Raman-active phonon modes [35].

Fig. 2 presents the typical time-domain TFISH and TKE results measured at 5 K, along with the corresponding Fourier-transformed frequency spectra. Under a weak THz field of 180 kV/cm, both measurements show negligible temporal responses after the main THz pulse (at time zero), indicating an insufficient driving force for the soft-mode coordinate. In contrast, the results at 540 kV/cm show well-defined multi-cycle oscillations following an initial rise in the response, superimposed on a non-oscillatory envelope lasting up to ~10 ps. Both frequency spectra exhibit a narrow peak near 1.3 THz and a broad peak near 1.9 THz, with their relative amplitudes differing between TFISH and TKE. These observations are qualitatively consistent with the previous study on $SrTiO_3$ [17]. Here we assign the 1.3 THz modes to the Raman-active $A_{1g}$ mode related to the AFD structural instability (anti-phase rotation of adjacent oxygen octahedra; see FIG. S6 [32]) and the 1.9 THz mode to an IR-active mode in the FE phase. Note that the THz-induced breaking of centrosymmetry has enabled simultaneous probing of both modes using either TFISH or TKE.

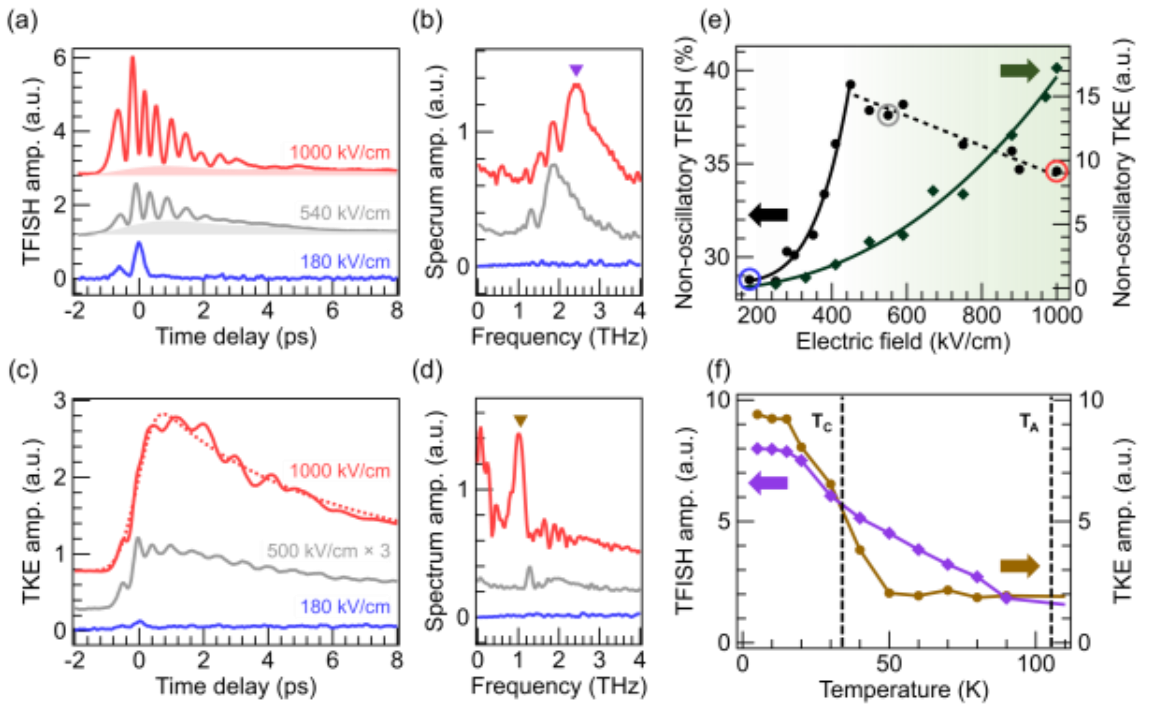

FIG. 2. Experimental observation of the hidden QPE phase. (a)─(d) TFISH (a) and TKE (c) signals measured at 5 K under three different THz field strengths and the corresponding Fourier-transformed frequency spectra (b and d). A significant oscillatory component can be observed in the time domain with increasing THz field, while the non-oscillatory component first grows and then decreases. The dashed lines in (c) denote the exponential function fitting to the non-oscillatory TKE component. (e) Comparison of the portions of the non-oscillatory TFISH and TKE components as a function of THz field. (f) Temperature-dependent mode strengths of the 2.4 THz TFISH mode and the 1.0 THz TKE mode.

We observed more striking results upon further intensifying the THz excitation. In the TFISH spectra measured at 1000 kV/cm, a prominent response mode emerges near the tail of the THz pulse spectrum, ~2.4 THz, closely aligning with the calculated second-order excitation energy of the FE soft mode [32]. Moreover, the non-oscillatory component appears to decrease compared to the 540 kV/cm case. This component, signifying the induced FE phase, was found to increase rapidly with THz field below 500 kV/cm but then decrease in a linear manner (Fig. 2E). These results indicate a critical THz field strength of 500 kV/cm, which is sufficient to drive the $Q_f$ into the second-order excited state. Consequently, competition between the QPE-FE and FE-hQPE phase transitions suppresses the non-oscillatory TFISH components. Meanwhile, the non-oscillatory component in the TKE spectra exhibits a monotonic, quadratic-in-field trend instead (Fig. 2E). This contrast confirms the origin of the hQPE phase as the restoration of inversion symmetry through coherent dipole moment transitions, whereas the dipole correlation consistently increases with the THz field (Fig. 1A).

In addition to the new 2.4 THz TFISH mode, the TKE spectra exhibit a dominant response mode near 1.0 THz at 1000 kV/cm, which we attribute to the Raman-active $A_{1g}$ phonon mode (note that the Raman- and IR-active modes now obey the rule of mutual exclusion in the hQPE phase). The temperature evolutions of these two modes offer insights into their coupled excitation behaviors. As shown in Fig. 2F (see full dataset in FIG. S9 [32]), the 2.4 THz mode begins to grow in magnitude below $T_A$, whereas the $A_{1g}$ mode becomes appreciable only near $T_C$. Since the AFD-type mode cannot couple directly to the THz field, its excitation depends on nonlinear coupling with the FE soft mode, such as linear-biquadratic coupling ($Q_{AFD}Q_f^2$), which necessitates a sufficient initial population of the latter mode. The recent study on $KTaO_3$ suggests that non-oscillatory TFISH signals may arise from field-induced dipole correlations, not necessarily a transient FE phase [31]. In our case, the multifaceted dynamical features observed, including nonmonotonic field evolution and mode coupling behaviors, provide unambiguous evidence for the proposed QPE-FE-hQPE transition pathway.

*Phonon behaviors of the hQPE phase*—We further address the phonon behaviors of the hQPE $SrTiO_3$. Fig. 3A and 3B present the full set of field-dependent TFISH and TKE frequency spectra, showing the systematic evolution of multiple phonon modes. Notably, the amplitude of the FE phase IR-active mode (green) followed a similar trend to the non-oscillatory TFISH components (Fig. 3C), again consistent with the rise-and-fall behavior of the FE phase. By contrast, the second-order excitation of the soft mode (violet) and the $A_{1g}$ mode (brown) increase monotonically, with the latter showing a highly nonlinear field dependence. As the THz field increases, the $A_{1g}$ mode, the FE phase IR-active mode, and the second-order excitation mode gradually harden, indicating a THz-induced deepening of the double-well potential in $SrTiO_3$ [26]. Accompanied by the breaking and restoration of centrosymmetry during the QPE-FE-hQPE phase transitions, the sensitivities of TFISH and TKE change simultaneously. The second-order excitation mode was prominent in TFISH and weak in TKE within the 400-700 kV/cm range, where the FE phase dominated and IR-active modes also became Raman-active. At higher THz fields, however, it was observed only in the TFISH spectra due to the rule of mutual exclusion. The rule was also verified for the $A_{1g}$ mode, which was prominent in TKE but barely observed in TFISH within the same field range.

The AFD ($A_{1g}$) mode softens with increasing THz field in the hQPE phase, further distinguishing itself from the FE soft modes (Fig. 3D). The AFD mode appears to strongly cooperate with the second-order excitation mode, as reflected by the large TKE response amplitude. This aligns with previous studies revealing a competition between the AFD and FE structural distortions [25]. Thus, the coupling to the AFD mode promotes the second-order excited state, thereby stabilizing the hQPE phase. Note that this coupling behavior has no counterparts in the equilibrium phase transitions of $SrTiO_3$ around $T_C$ or $T_A$. We also note two additional TKE modes near 0.1 THz and 0.25 THz occurring at high THz fields. These Raman-active modes grow highly nonlinearly with THz field and show no frequency shifts. We point out that these modes have not been observed in equilibrium states of $SrTiO_3$, and possibly arise from the mixing of multiple AFD modes along different directions.

In light of all experimental results, we emphasize that the hQPE phase in $SrTiO_3$ is fundamentally distinct from its QPE ground state. The QPE phase lacks the long-range correlation of the soft-mode dipoles due to zero-point energy, whereas the hQPE phase loses the dipole moment collectively in the presence of an established FE order due to the multiply excited quantum states. Therefore, hQPE $SrTiO_3$ represents a unique non-equilibrium state with intricate lattice dynamics. We expect that this state can be tuned by adjusting not only the field strength but also the spectral content of THz pump pulses. Also, the substantially enhanced AFD mode susceptibility in the hQPE phase is a remarkable property, which could have important implications for THz-induced dynamic multiferroicity in $SrTiO_3$.

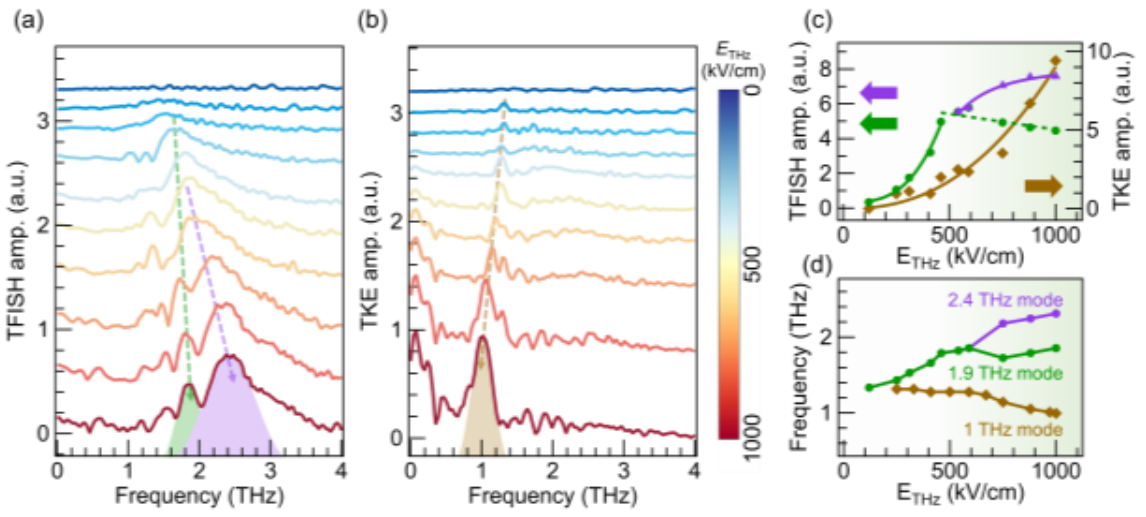

FIG. 3. Phonon behaviors of the hQPE phase. (a, b) Fourier-transformed spectra of the TFISH (a) and TKE (b) signals measured at 5 K under a sequence of THz field strengths. The resonance response peaks of the first-order, second-order excitation of the FE soft mode and the AFD mode are shadowed in green, violet and brown, respectively. (c) Power or exponential function fittings to the three designated modes. The FE phase IR-active mode closely follows the evolution of the induced FE phase portion. (d) Field evolutions of the phonon mode frequencies. The FE phase IR-active mode and the second-order FE soft mode hardens while the AFD mode softens with increasing THz field.

*First-principles calculations*—To gain more insights into the hQPE phase transition, we performed dynamical simulations using the time-dependent Langevin-Schrödinger equation (TDSE) and *ab initio* molecular dynamics (AIMD) methods. The TDSE was solved with the effective potentials for interactions between the soft mode ($Q_f$), AFD mode ($Q_a$), and lattice strain ($Q_c$), as evaluated from density functional theory (DFT) total energy calculations [26]. The THz excitation was incorporated via a coupling term in the TDSE, representing the dipole interaction between the external electric field and the effective charge of the soft mode. Single-cycle resonant THz pulses ($\omega$ = 0.5 THz) with varying peak field strengths were applied, and the resulting temporal evolutions of $Q_f$ were examined, as shown in Fig. 4A. At a low field strength (160 kV/cm), $Q_f$ oscillates for a few picoseconds upon THz excitation, but eventually returns to zero due to dissipation. At a moderate field strength (450 kV/cm), $Q_f$ oscillates similarly but stabilizes at a non-zero value. At a high field strength (1130 kV/cm), $Q_f$ exhibits both low- and high-frequency oscillations, with the latter attributed to the second-order excitation of the FE soft mode, before decaying to zero. The averaged values of $Q_f$ were calculated to determine their THz field dependence (Fig. 4B). A non-zero shifted value ($\Delta$) is observed only within an intermediate field range of 280–450 kV/cm, with the lower and upper critical fields corresponding to the QPE-FE and FE-hQPE phase transitions, respectively. The TDSE frequency spectra show that, along with the resonantly excited soft mode near 0.5 THz, a significant mode emerges near 2.8 THz at high field strengths (Fig. 4C), consistent with the calculated second-order eigenfrequency (Fig. 1B). Moreover, the excitation strengths of the second-order soft mode and the AFD mode appear to be correlated in the hQPE phase (Fig. 4B inset), implying a cooperative relationship between them. These TDSE simulation results are corroborated by the TFISH and TKE observations.

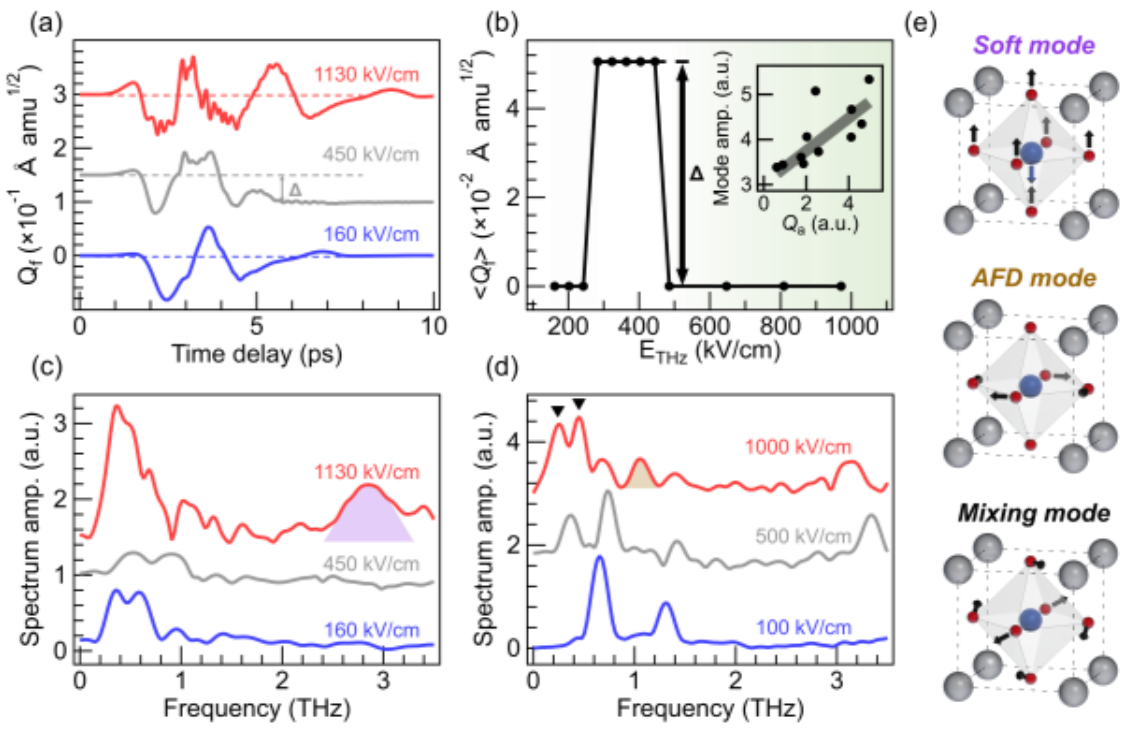

FIG. 4. THz-induced dynamics in SrTiO3 simulated by the TDSE and AIMD methods. (a) Temporal evolution of the soft-mode coordinate $Q_f$ in the TDSE simulations under THz field strengths of 160, 450, and 1130 kV/cm. (b) The shifted time-average $Q_f$ values (calculated after the initial fluctuations during the THz pulse) as a function of THz field. The inset shows the correlation between the AFD mode and the second-order excitation of the FE soft mode. (c) Fourier-transformed spectra of $Q_f$ under the same field strengths in (a). (d) Fourier-transformed spectra of the oxygen atom dynamics in the AIMD simulations. THz excitation conditions of 100 kV/cm, 500 kV/cm and 1000 kV/cm were evaluated by setting the initial atomic displacements according to the corresponding maximum $Q_f$ values in the TDSE simulations. (e) Schematics of the FE soft mode and the AFD modes corresponding to the eigenmodes in (c) and (d).

The AIMD results further clarified the field evolution of the AFD mode characteristics. We simulated the THz excitation by setting initial atomic displacements according to the THz-induced initial $Q_f$ values from the TDSE results. Fig. 4D presents the Fourier-transformed spectra of the oxygen rotational motions. At a low THz field (100 kV/cm), two well-defined modes are observed at ~0.65 THz and ~1.2 THz, both of which split as the THz field increases. By analyzing the relative phases of the oxygen ion motions [32], we attribute the 1.2 THz mode to the AFD $A_{1g}$ mode with pure in-plane rotation of oxygen ions (Fig. 4E). At high THz fields (1000 kV/cm), the $A_{1g}$ mode softens to ~1.0 THz, and two new mixed modes appear in the low-frequency range. The predicted softening of the $A_{1g}$ mode and emergence of new mixed modes are consistent with the TKE observations (Fig. 3B). The mixed modes involve multiple AFD rotations along the *x*, *y* and *z* directions, with slight phase differences between these rotations. Notably, these AFD-type modes are unique to the hQPE phase, potentially modifying the electronic properties of $SrTiO_3$.

*Discussion and summary*—Based on the evident ultrafast THz spectroscopy data assisted by first-principles calculations, we have unambiguously demonstrated the generation of a new and hidden quantum state in $SrTiO_3$ along the QPE-FE-hQPE phase transition path and identified the mechanism of high-order excitation of the soft phonon mode driven by intense THz pulses. These findings represent a novel strategy for ultrafast coherent manipulation of materials, which can be directly extended to other ferroelectrics where the occurrence of polarization is inherently linked with soft-mode nonlinearities. The precise control and switching of polarization states on ultrafast timescales could enable emerging applications in data processing, computing and optoelectronic devices. Beyond polar materials, phonon nonlinearities are also ubiquitous in correlated systems, such as high-temperature superconductors and magnetic materials; manipulating their dynamics to elicit temporal functional responses also constitutes a research frontier with significant technological prospects. We further note that recent developments in extremely strong THz sources [36] and advanced THz optics, including filters, waveplates, and cavities [37,38], can considerably enhance the efficacy of high-order phonon excitations and provide additional tuning capabilities, such as selective excitation of specific phonon eigenmodes. Altogether, we anticipate that our method will offer dynamic access to unprecedented structural transitions and uncover exotic physical phenomena in quantum materials that cannot be achieved through conventional methods.

The authors would like to thank Prof. Shi Liu for useful discussions. A portion of this work was carried out at the Synergetic Extreme Condition User Facility (SECUF, https://cstr.cn/31123.02.SECUF). This work was financially supported by The Basic Science Center Project of National Natural Science Foundation of China (NSFC) grant No. 52388201, NSFC grant No. 12474087 and No. 11974414, National Key Basic Research Program of China grant No. 2020YFA0309100, Beijing Natural Science Foundation grant No. JQ24011 and No. Z240008, National Research Foundation of Korea (NRF) grant funded by the Korea government (MSIT) (No. RS-2024-00333664 and No. RS-2023-00218180) and the Ministry of Science and ICT (No. 2022M3H4A1A04074153).

Q.L. and D.S. conceived and supervised this study; W.L., X.W. and J.L. conducted ultrafast THz spectroscopies; H.K. and D.S. performed first-principles calculations; S.L., J.-M.L., J.-F.L., A.R. and C.-W.N. discussed the results. W.L., D.S. and Q.L. co-wrote the manuscript with input from all the co-authors.

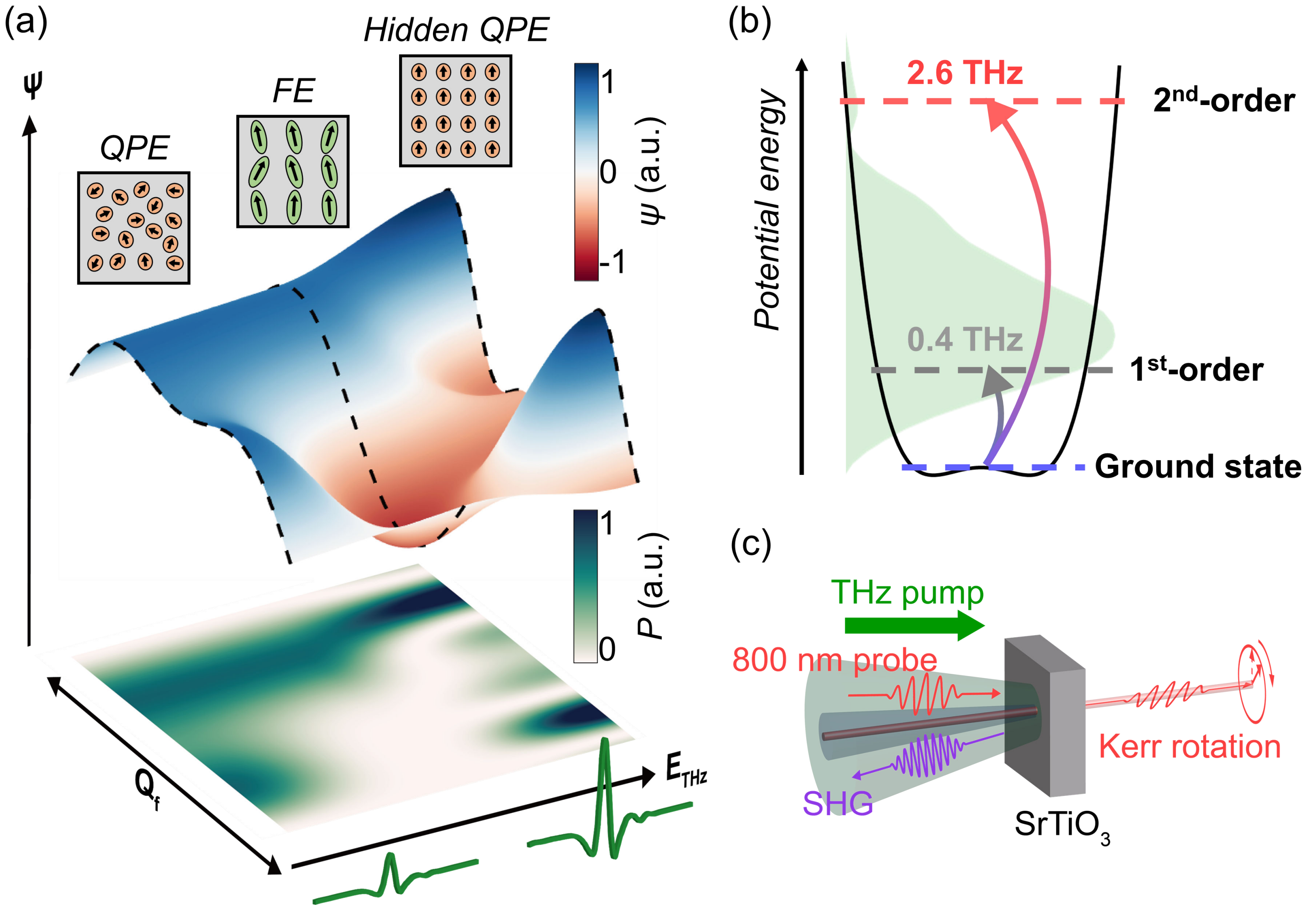

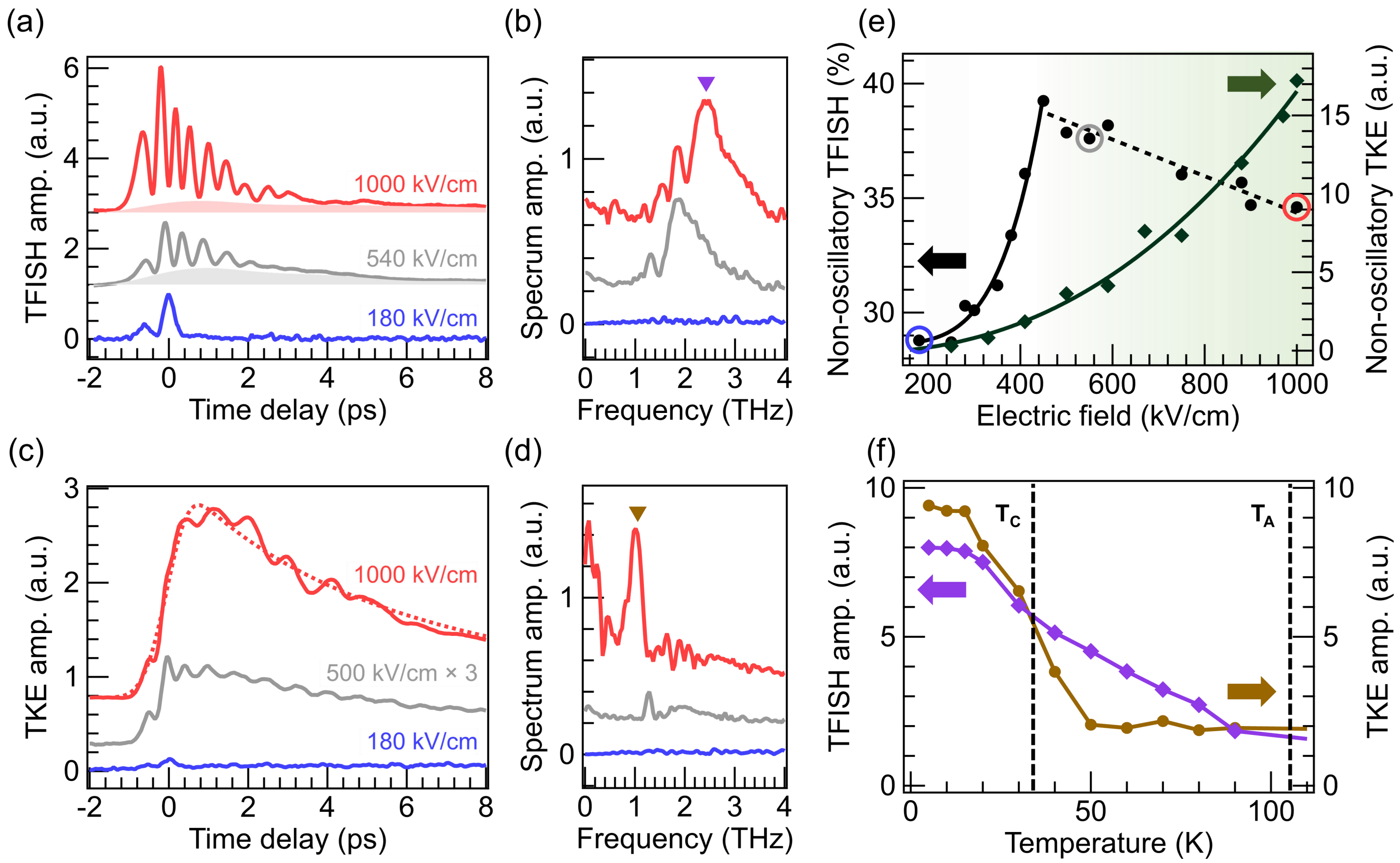

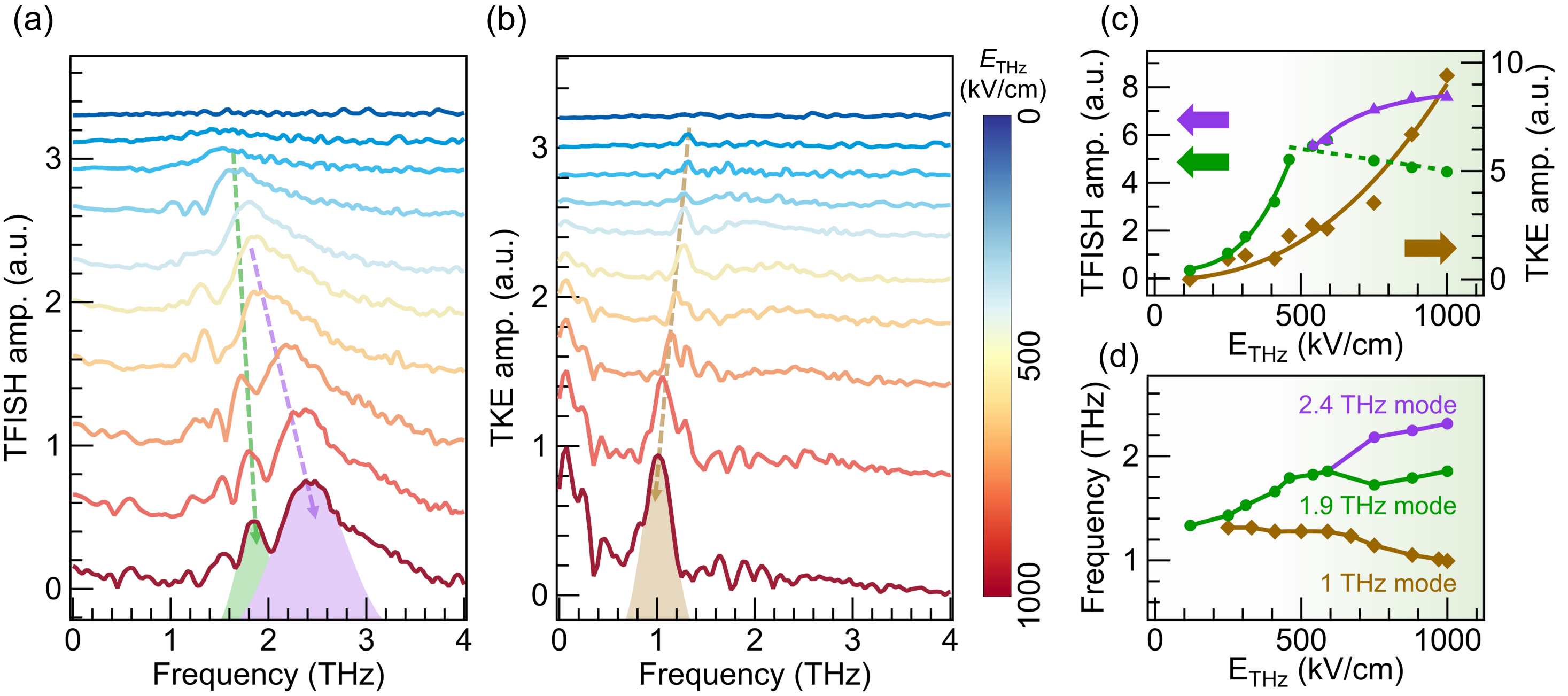

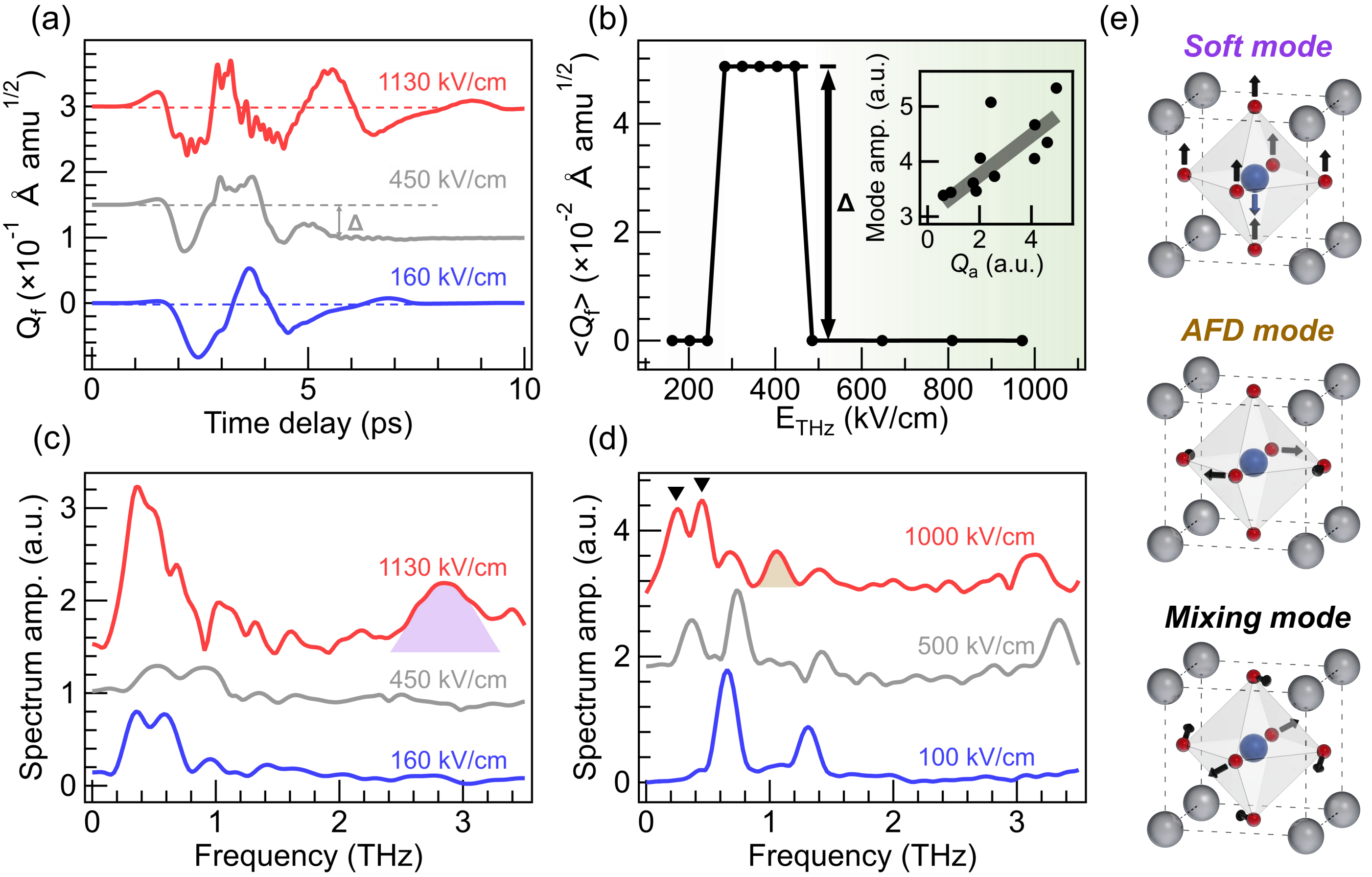

# Supplementary Materials for:
# A Hidden Quantum Paraelectric Phase in $SrTiO_3$ Induced by Terahertz Field

This supplemental material provides additional details on the research methods, the analysis methods of the experiment data, the first-principles calculation results, and the ultrafast spectroscopy results.

## A. Details on the First-principles Calculation methods

To investigate the real-time dynamics in the quantum paraelectric $SrTiO_3$, we employed the time-dependent lattice Langevin-Schrodinger equation (TDSE) as follows:

$$i\hbar\frac{d}{dt}\psi = \widehat{H}_{2D}\psi + \gamma(\hat{S} - <\hat{S}>) \tag{S1}$$

where $\gamma = 1.2$ THz and $\hat{S}$ are the dissipation rate and wavefunction phase. The fractional dissipation effect for lattice wave function is described by Langevin-Schrodinger formalism. The Hamiltonian for TDSE is given as follows:

$$\begin{aligned}\widehat{H_{2\mathrm{D}}}\left[\widehat{Q_f},\widehat{Q_c},Q_a(t),t\right] = {}& \widehat{P_f^2}/2M_f \\ &+\widehat{P_c^2}/2M_c \\ &+\widehat{V^{\mathrm{f,c}}}\left[\widehat{Q_f},\widehat{Q_c}\right] \\ &+\widehat{V^{\mathrm{f,c,a}}}\left[\widehat{Q_f},\widehat{Q_c},Q_a(t)\right] \\ &+E_{ext}(t)Z_f\widehat{Q_f}\end{aligned} \tag{S2}$$

In this Hamiltonian, the soft mode ($Q_\mathrm{f}$), lattice strain ($Q_\mathrm{c}$), antiferrodistortive mode ($Q_\mathrm{a}$), and their nonlinear interactions are considered. $\widehat{P_f^2}/2M_f$ , $\widehat{P_c^2}/2M_c$ , and $E_{ext}(t)Z_f\widehat{Q_f}$ are kinetic terms for the soft mode, the lattice strain, and the dipole interaction between the external electric field and the soft mode via mode effective charge, respectively. $Q_\mathrm{f}$ and $Q_\mathrm{c}$ are described in 2-dimensional (2D) wavefunctions and $Q_\mathrm{a}$ is treated in classical Newtonian dynamics. To describe the 2D lattice wavefunction we used the 200×200 numerical grids. The effective interaction potentials ($\widehat{V^{\mathrm{f,c}}}\left[\widehat{Q_f},\widehat{Q_c}\right] + \widehat{V^{\mathrm{f,c,a}}}\left[\widehat{Q_f},\widehat{Q_c},Q_a\right]$) between these three components are evaluated from density functional theory (DFT) calculation using Quantum Espresso package as follows:

$$\begin{aligned}V^{\mathrm{f,c}} = {}& \sum_{i=1}^{12} k_{f,i}\,\widehat{Q_f^{2\imath}} \\ &+\sum_{j=2}^{10} k_{c,j}\,\widehat{Q_c^{\jmath}} \\ &+\sum_{i=1}^{12}\sum_{j=1}^{10} k_{fc,i,j}\,\widehat{Q_c^{\jmath}}\,\widehat{Q_f^{2\imath}}\end{aligned} \tag{S3}$$

$$\begin{aligned}V_a = {}& \sum_{i=1}^{10} k_{a,i}\,Q_a^i \\ &+\sum_{i=1}^{10}\sum_{j=1}^{12} k_{af,i,j}\,Q_a^i\langle Q_f\rangle^{2j} \\ &+\sum_{i=1}^{10}\sum_{j=1}^{10} k_{ac,i,j}\,Q_a^i\langle Q_c\rangle^{j}\end{aligned} \tag{S4}$$

$$\begin{aligned}V^{f,c,a}\left[\widehat{Q_f},\widehat{Q_c},t\right] = {}& \sum_{i=1}^{10}\sum_{j=1}^{12} k_{af,i,j}\,Q_a^i(t)\widehat{Q_f^{2\jmath}} \\ &+\sum_{i=1}^{10}\sum_{j=1}^{10} k_{ac,i,j}\,Q_a^i(t)\widehat{Q_c^{\jmath}}\end{aligned} \tag{S5}$$

The same parameters for TDSE and effective interaction potentials evaluated by DFT calculations are employed from the previous report [1]. Note that the TDSE and Newtonian equation are solved at the same time steps and their information $<\widehat{Q_f}(t)>$, $<\widehat{Q_c}(t)>$, and $Q_a(t)$ are updated for interaction potentials.

The electron-electron exchange and correlation potentials are described through the Perdew-Berke-Ernzerhof functional. We consider the $\sqrt{2}\times\sqrt{2}\times 2$ unit cell for the low-temperature tetragonal phase of $SrTiO_3$ and we sample the Brillouin zone with a $6\times 6\times 4$ $k$-point grid. The projector augmented wave (PAW) method was considered to describe the core level of atomic orbitals, and the plane wave basis set with energy cut-off 70 Ry was employed.

We performed ab initio molecular dynamics (AIMD) simulations to investigate the complex motion of oxygen in a non-equilibrium state. Because the previous TDSE is restricted to three components (soft mode, lattice, and AFD mode), capturing the nonlinear phonon interaction with the other modes with the highly excited soft mode is hard. We consider the $2\times 2\times 2$ unit cell for the tetragonal phase of $SrTiO_3$, which allows the AFD modes along the x, y, and z-axis and a total of 120 phonon modes. We sample the Brillouin zone with a $2\times 2\times 2$ $k$-point grid. The Perdew- Zunger local density approximation functional and PAW method are employed with the plane wave basis set with energy cut-off 70 Ry for electron-electron exchange-correlation energy and core levels, respectively. To mimic the THz pump-induced dynamics in AIMD simulation, we initially displaced the atomic geometry toward the soft mode eigenvector direction with its maximum amplitude achieved from TDSE simulation. With this initial condition, we performed the AIMD simulation without the temperature control for the non-equilibrium dynamics, and the time propagation up to 10 picoseconds with the 5fs time step was considered. To check the phonon softening of AFD mode as an experiment, we introduced the strained condition on the lattice (a=3.86 Ang and c/a=1) to get the 1.2 THz frequency for AFD mode.

| Excitation state | Energy (eV) | Frequency (THz) |
|---|---|---|
| Ground state | 0 | 0 |
| 1 | 0.0016 | 0.387 |
| 2 | 0.0107 | 2.589 |
| 3 | 0.0168 | 4.065 |
| 4 | 0.0243 | 5.879 |

TABLE SI. The TDSE simulation results of different excitation states of the soft mode in $SrTiO_3$, along with the corresponding phonon frequencies.

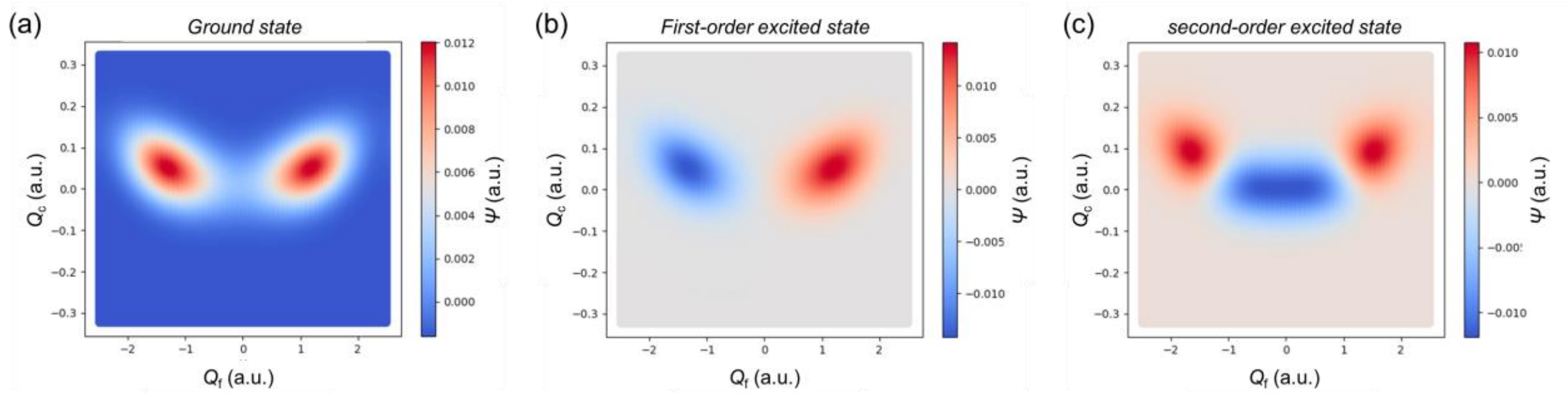

FIG. S1. First-principles calculation results of the wave function $\Psi$ in $SrTiO_3$. The distributions of $\Psi$ are plotted as a function of the soft mode coordinate ($Q_f$) and the lattice strain ($Q_c$). Three different conditions are considered for (a) the ground state, (b) the first-order excited state and (c) the second-order excited state of the soft mode.

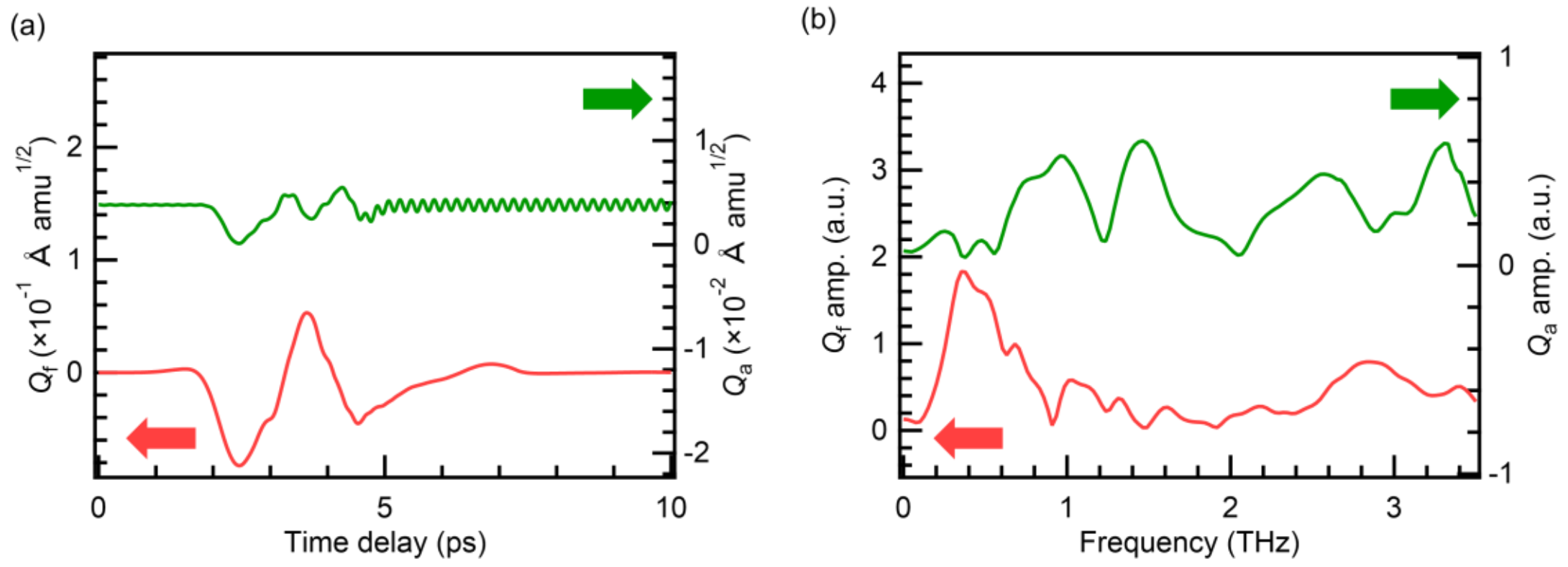

FIG. S2. Determination of phonon modes in the TDSE simulations. (a) The time-domain evolution of $Q_f$ and $Q_a$. (b) The corresponding Fourier-transformed spectra.

### B. Details on the Ultrafast Experiment Setup

The samples were commercial $[001]_c$-oriented $5\times5\times0.5$ mm$^3$ $SrTiO_3$ single crystals and were double side-polished for the transmission geometry measurements.

The TFISH and TKE measurements were performed with a free-space optical system. The 35 fs laser pulses were generated from a Ti:sapphire femtosecond laser amplifier (Spitfire Ace, Spectra-Physics) with a repetition rate of 1 kHz, a central wavelength of 800 nm, and a single-pulse energy of 7 mJ. The laser beam was separated into the THz generation line (95%) and the probe line (5%). For the generation line, THz radiation was generated from a $LiNbO_3$ single crystal by the tilted-wavefront technique [2], and focused onto the sample in normal incidence by a pair of off-axis parabolic mirrors. The direction of $SrTiO_3$ was adjusted to align the in-plane $[100]_c$ crystal axis with the THz field. The temporal profile of the THz field was measured by the electro-optic sampling (EOS) method using a GaP single crystal. For the probe line, the remaining weak laser output was reflected by several mirrors to be collinear with the THz radiation and focused onto the sample by a convex lens (FIG. S3).

In the TFISH measurements, the THz-induced SHG signals were collected in a near-backscattering geometry by a photomultiplier tube. A half-waveplate and a polarizer were used to control the polarization direction of the probe and SHG light, respectively. The polarization of the probe light and the SHG light were set to be parallel to the THz pulses. In the TKE measurements, the probe beam was polarized at 45° relative to the THz field. The transmitted light was separated by a quarter-waveplate and a Wollaston prism, and collected with a balanced detector. The THz-induced Kerr effect was then measured through the depolarization of the probe light field. In both measurements, a mechanical chopper was used to modulate the THz generation beam, and the recorded signals were demodulated from the original stream using a lock-in amplifier.

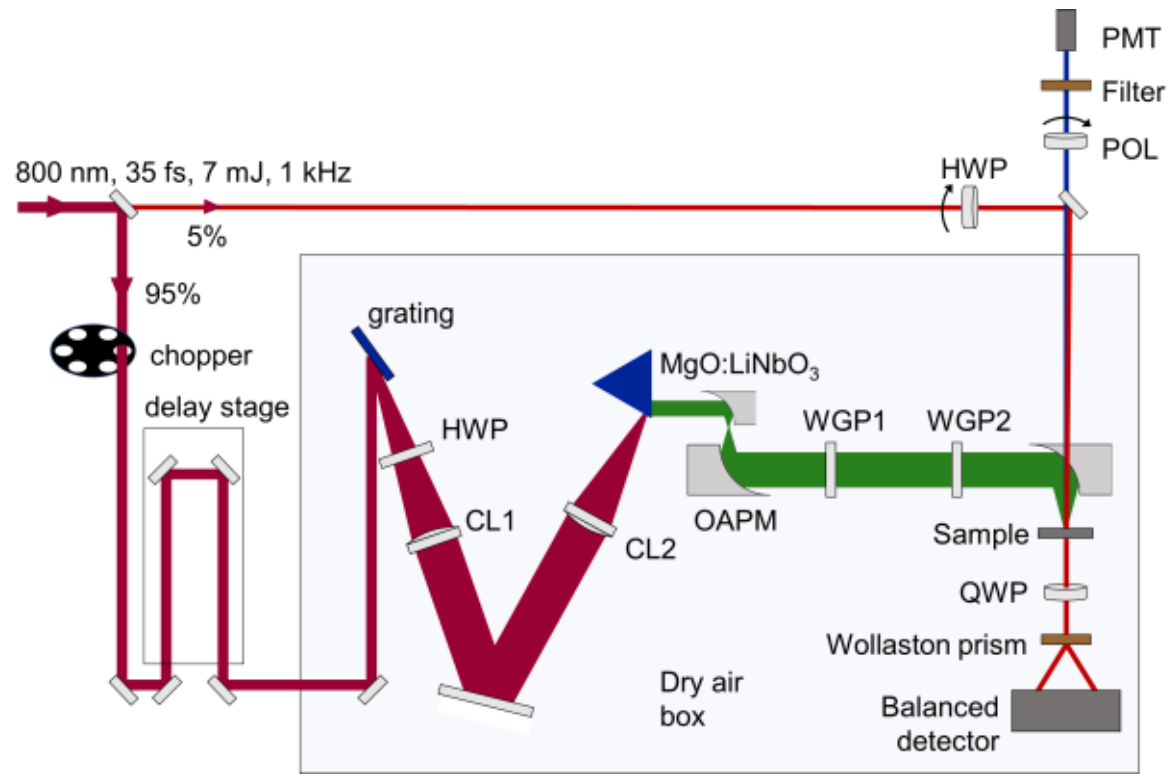

FIG. S3. Illustration of the experiment setup. The 800 nm, 35 fs laser with a single-pulse energy of 7 mJ was separated into the THz generation line (95%) and the probe line (5%).THz radiation was generated from a $LiNbO_3$ single crystal by the tilted-wavefront technique. For the probe line. A dry air box was used to minimize the influence of water vapor. The TFISH signal and the TKE signal were collected with a photomultiplier and a balanced detector, respectively. HWP: half-wave plate. CL: plano-convex cylindrical lens. OAPM: off-axis parabolic mirror. WGP: terahertz wire grid polarizer. QWP: quarter-wave plate. POL: polarizer. PMT: photomultiplier tube.

The strength of the THz radiation was adjusted with a pair of polarizers, and the maximum field strength was estimated to be 1000 kV/cm. For the temperature-dependent measurements, a cryostat with a TPX window was used to minimize the loss of THz field strength. All measurements were performed in a box filled with flowing dry air, where the moisture was controlled below 5% to eliminate the influence of water vapor on the measured THz spectra.

### C. Determination of the phonon modes in first-principles calculations

In the TDSE simulations, soft mode ($Q_f$), AFD mode ($Q_a$) and lattice strain ($Q_c$) were considered. We tracked the evolution of $Q_f$ and $Q_c$ by calculating the time-domain evolution of the mode coordinates and the corresponding Fourier-transformed spectra. As shown in FIG. S2, the mode amplitudes of both modes were summarized by calculating the integral area of certain peaks in the Fourier-transformed spectra.

In the AIMD simulations, the motions of the oxygen atoms in the model (2 × 2 × 2 unit cell) are tracked in the form of x, y and z movements. We first numbered the oxygens at different sites as shown in FIG. S4. The Fourier transformations were then performed and 6 peaks with large amplitudes were selected as shown in FIG. S5. The amplitudes and phases of all oxygen atoms at these frequencies are summarized in FIG. S7 and FIG. S8 for 100 kV/cm and 1000 kV/cm, respectively. The 0.75 THz mode is characterized by the out-of-phase motion of adjacent oxygen atoms in the x-y plane, indicating the AFD rotation. It is mixed with the out-of-phase motion of adjacent oxygen atoms in the z-direction, indicating another AFD rotation in the y-z plane. With the phase information, we reconstruct the phonon mode into an illustration in FIG. S6 (mixing mode *left*). Similar assignments can be done to the other 5 modes and was summarized in TABLE. SII.

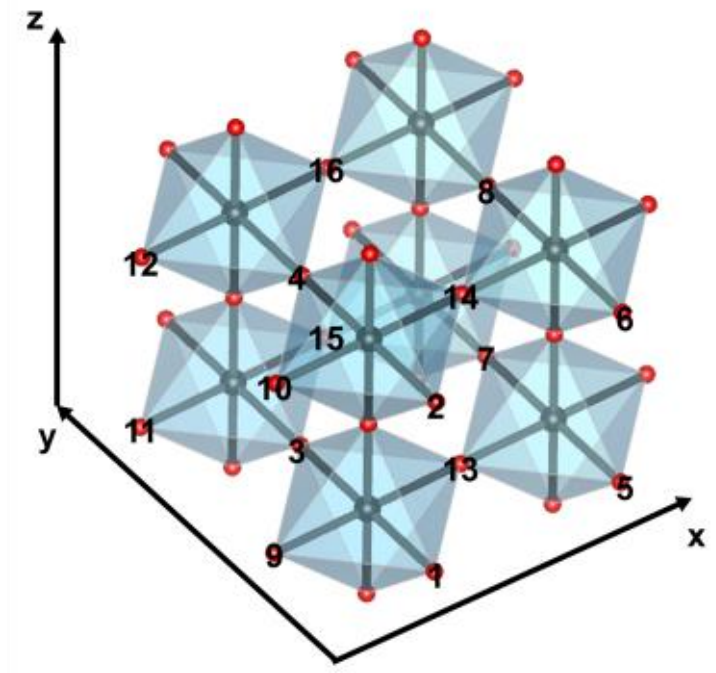

FIG. S4. Illustration of the atomic model in the AIMD simulation and the numbering of oxygen atoms.

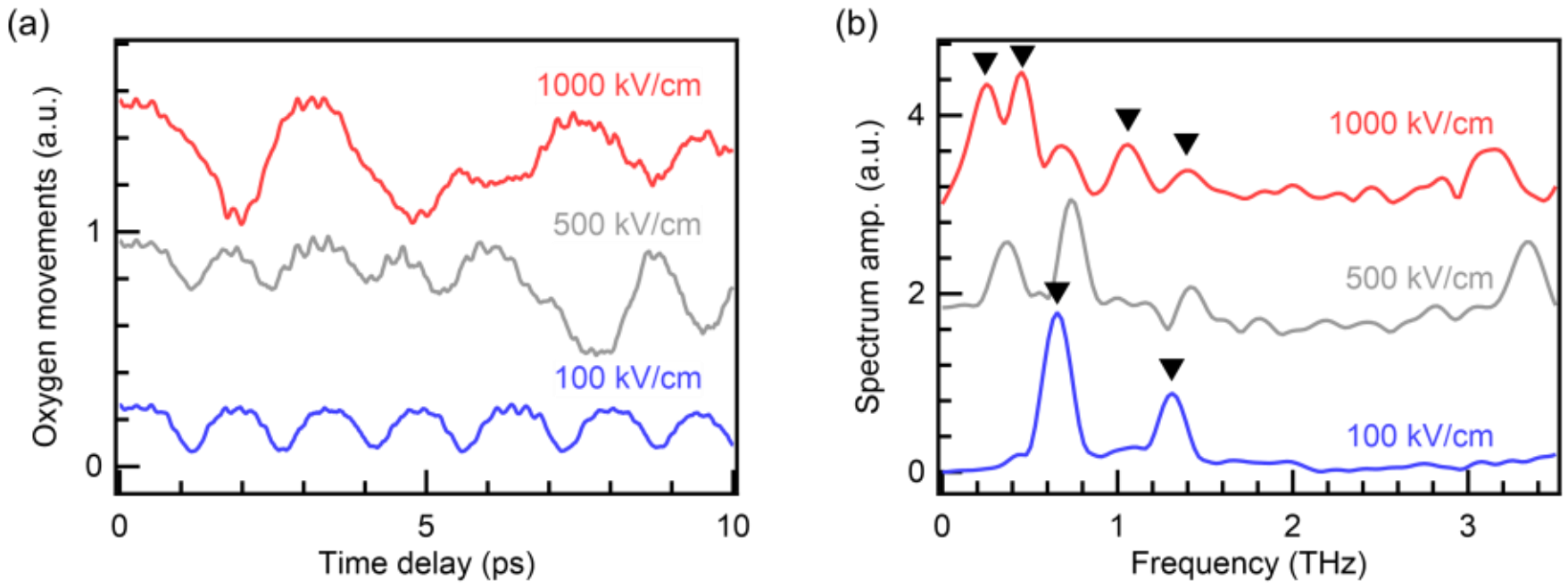

FIG. S5. AIMD simulation results at 100 kV/cm, 500 kV/cm and 1000 kV/cm. (a) Time-domain evolution of oxygen motions. (b) The corresponding Fourier transformed spectra.

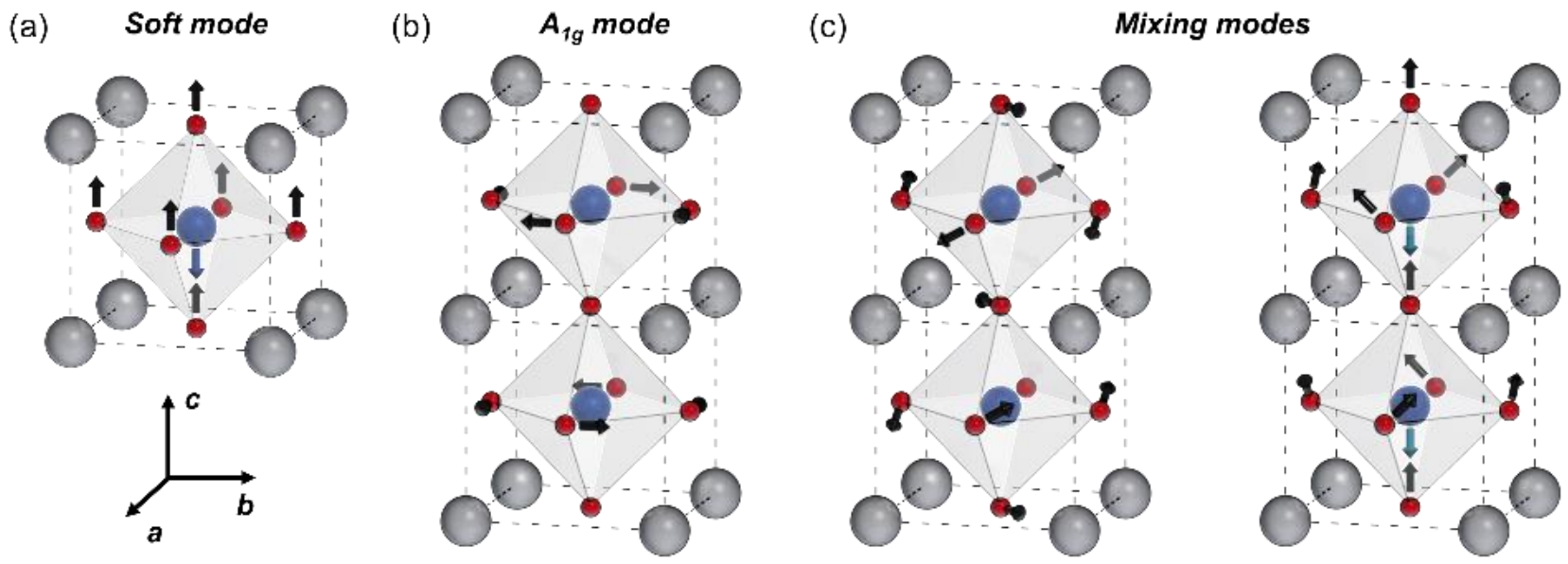

FIG. S6. Illustration of the different phonon modes in the terahertz band in the hQPE phase of $SrTiO_3$. (a) The soft mode (Slater mode) of $SrTiO_3$. The cation (Ti) and anion (O) move in opposite directions along the out-of-plane direction. (b) The Raman-active $A_{1g}$ mode. The tetragonal unit cell of low-temperature phase of $SrTiO_3$ is characterized by the rotations of two oxygen octahedra about the c-axis in opposite directions known as antiferrodistortive rotation. The phonon modes which involve the same rotations are called AFD modes. (c) New mixing phonon modes identical to the hQPE phase of $SrTiO_3$ as calculated by AIMD simulations. The left one can be regarded as a mixture of AFD modes about both the c- and a- axes. The right is a mixture of the c-axis AFD mode and the soft mode.

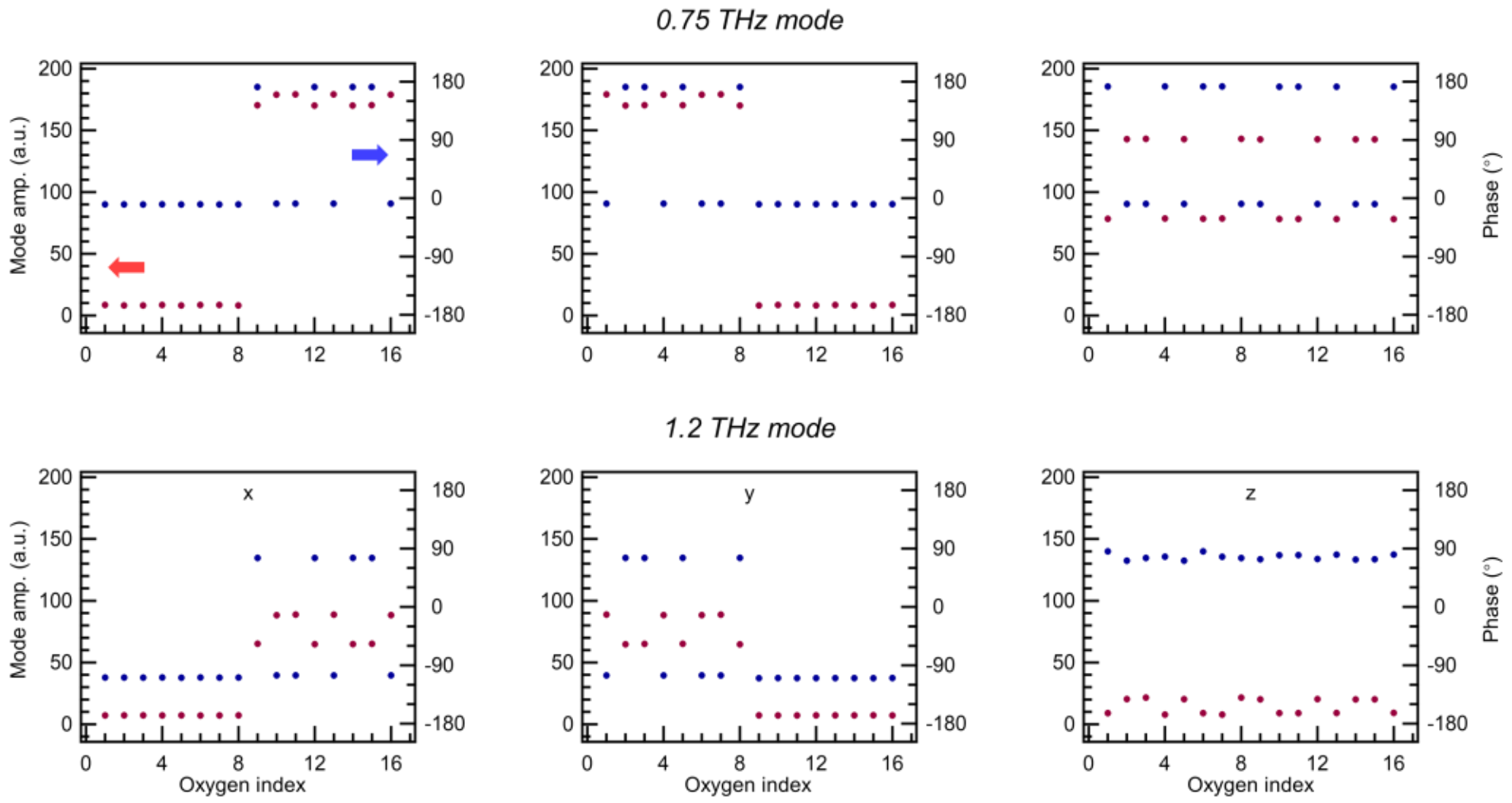

FIG. S7. Full data set of the Fourier-transformed results of the AIMD simulation at 100 kV/cm. The motions of all oxygen atoms are recorded in the form of the amplitudes (red dots) and phases (blue dots) of certain modes.

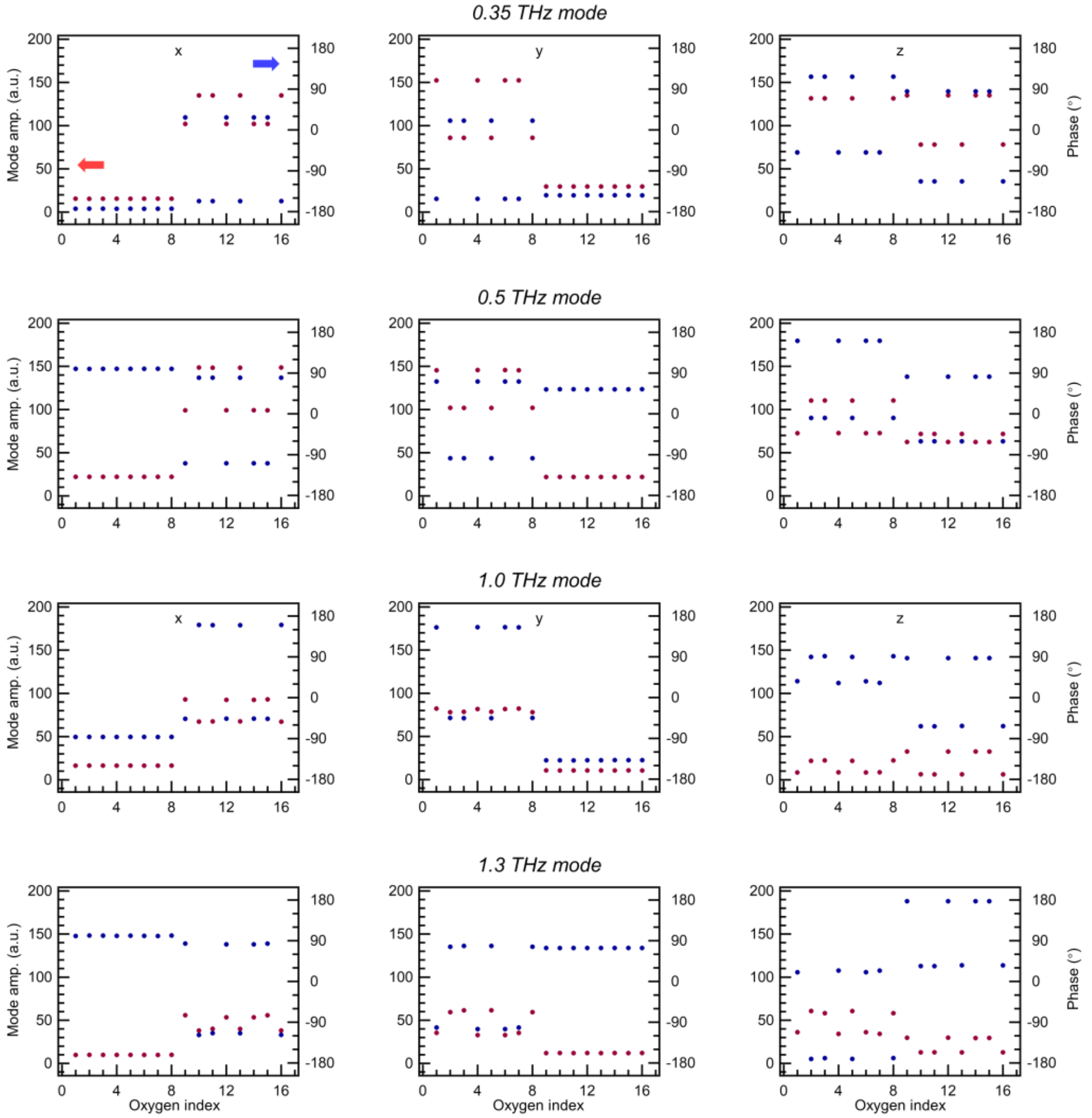

FIG. S8. Full data set of the Fourier-transformed results of the AIMD simulation at 1000 kV/cm.

| Field strength (kV/cm) | 100 | | 1000 | | | |
|---|---|---|---|---|---|---|
| Frequency Assignment | 0.75 THz mixing | 1.2 THz AFD | 0.35 THz mixing | 0.5 THz mixing | 1.0 THz AFD | 1.3 THz AFD |

TABLE. SII. Assignment of phonon modes according to the AIMD simulation.

### E. Calculation of the Fourier-transformed spectra

For all the frequency-domain TFISH and TKE spectra, and the simulation results in TDSE and AIMD, the numerical time derivatives of the original time-domain data were calculated first to exclude the influence of non-oscillatory components. Fourier transformations were then performed by applying a Hanning window function and an appropriate zero-padding. The analysis of the non-oscillatory components in the main text Fig. 2 was conducted in a different way. We performed fast Fourier transformation directly on the time-domain data, and selected the zero-point (0 THz) component as the extent of non-oscillatory component. Next, an artificial spectrum with the exact values (amplitude and phase) at the zero point and zero values at other frequencies was constructed. Inverse Fourier transformation was then performed to simulate the time-domain evolution of this non-oscillatory component.

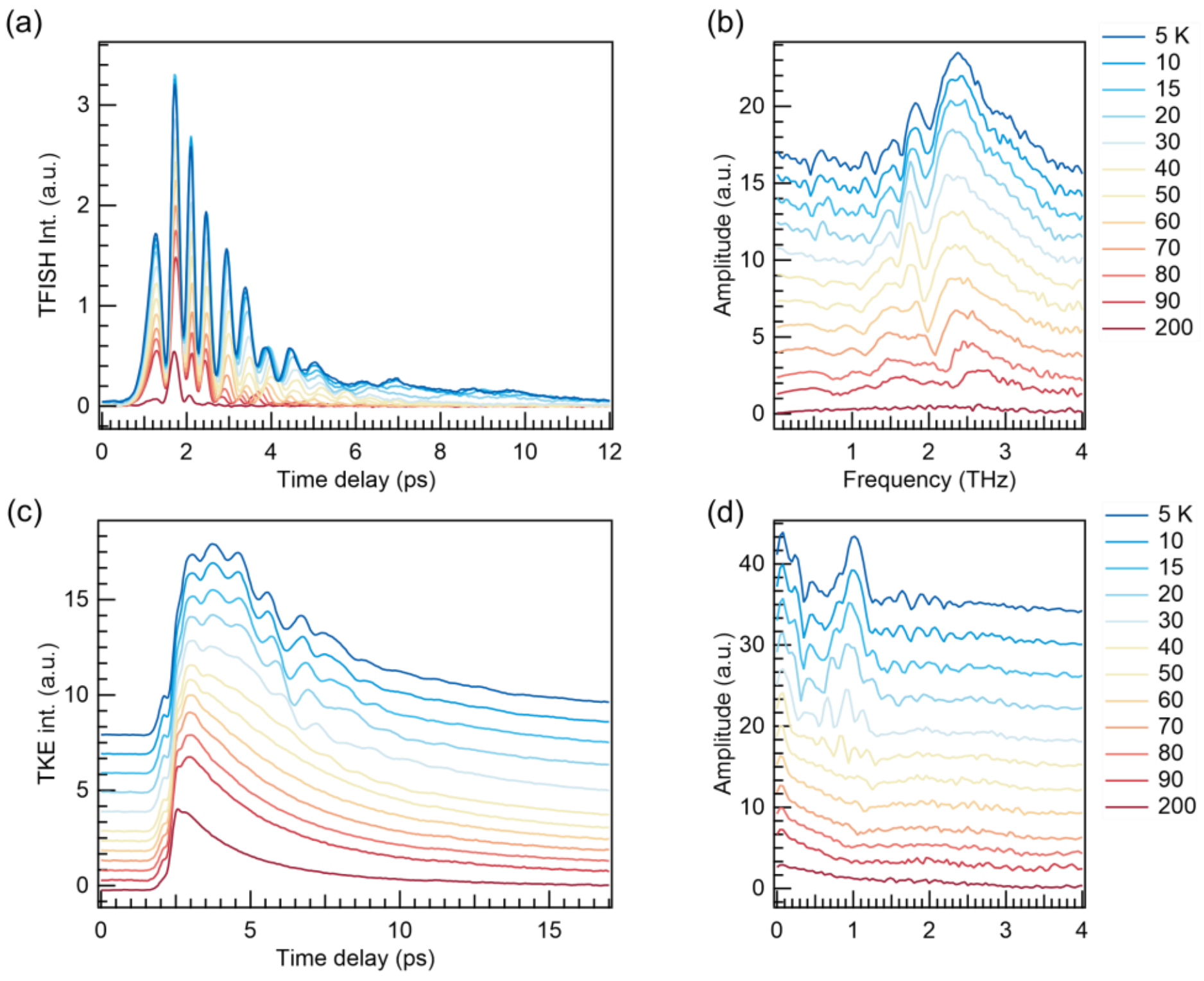

FIG. S9. Full dataset of the TFISH and TKE measurements under varying temperatures at a fixed THz field strength of 1000 kV/cm. (a) THz-pump, SHG-probe experiment results and (b) the corresponding Fourier transformed spectra. (c) THz-pump, Kerr-probe experiment results and (d) the corresponding Fourier-transformed spectra.

### D. Measurement of the THz field strengths

The field strengths are critical in the accesses to the FE and hQPE phases in $SrTiO_3$ as comprehensively discussed in the Main Text. To precisely measure the field strengths of the generated THz pulses, we quantitatively performed electro-optic sampling (EOS) and applied a THz camera, as shown in FIG. S10. For EOS measurements, a pair of terahertz wire grid polarizers were used to suppress the THz field strength to 3.6% of the origin value. Four high-resistance wafers (field transmission $t_{Si}$, of 70%) were also inserted to protect the detection crystal from saturation. EOS was then performed with a GaP single crystal with thickness $L$=0.5 mm. At the peak of the THz field (2 ps), the relative intensity change collected by the balanced detector is $\Delta I/I = 3.6\%$. The THz field strength can then be calculated by the following equation:

$$\arcsin\left(\frac{\Delta I}{I}\right) = \frac{2\pi n_0^3 r_{41} t_{GaP} t_{Si}^4 E_{THz} L}{\lambda} \quad \text{(S6)}$$

where $n_0 = 3.32$ is the refractive index of the detection crystal GaP at the THz band, $r_{41} = 0.88$ pm/V is the electro-optic coefficient, $t_{GaP} = 2/(1+n_0) = 0.46$ is the Fresnel transmission coefficient, and $\lambda_0 = 800\ nm$ is the wavelength of the probing light. From the equation the peak intensity of the THz pulses is calculated to be 970 kV/cm.

We next used a THz power meter (Ophir 3A-P-THz) to measure the maximum power to be 6.4 mW at the sample position, corresponding to a single THz pulse energy of $W_{THz} = 6.4\ \mu J$. At the same position, a teraherz camera (Swiss Terahertz, S2X) was set to capture the image of the THz spot. The spot diameters along the horizontal and the vertical directions were $a = 0.53\ mm$ and $b = 0.60\ mm$, respectively. The THz field strength can then be derived from the following equation [3]:

$$W_{THz} = \frac{\pi}{8}\epsilon_0 cab \int E_{THz}^2(t)dt \quad \text{(S7)}$$

where $c$ and $\epsilon_0$ are the light speed and the dielectric constant in vacuum. The THz field strength was calculated to be $E_{THz} = 1250\ kV/cm$. Note that the measurement by THz camera produced larger errors due to the inaccuracy of the thermometer-based device. We therefore conclude that the obtained peak intensity of the THz field is around 1000 kV/cm.

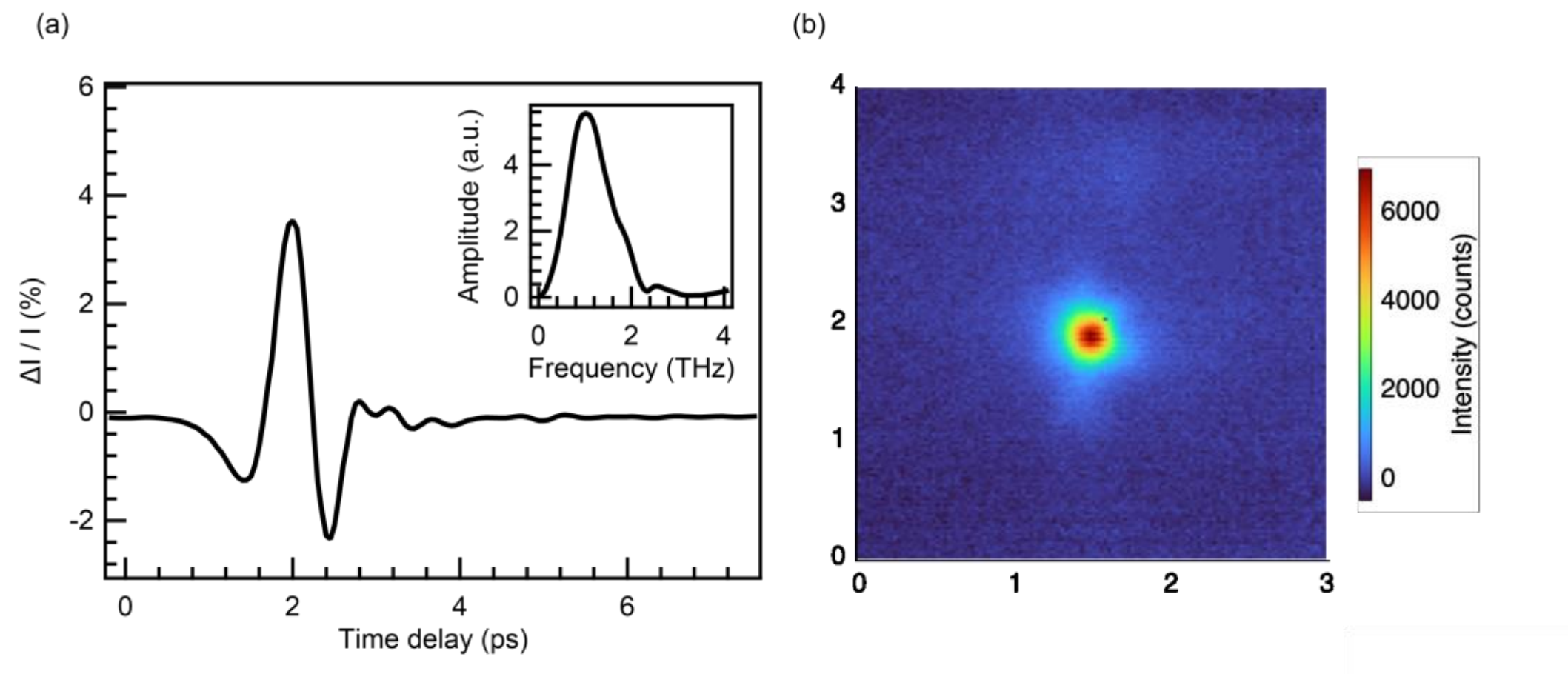

FIG. S10 Measurement results of the THz field strengths. (a) Quantitative electro-optic sampling result together with the corresponding Fourier transformed spectra (inset). (b) Image of the THz spot.

### F. Consideration of the response time

Generally, light-induced phase transition of quantum materials can happen in a few picoseconds. In the particular case of $SrTiO_3$, the transition into the FE phase is evident by the large amplitude of the excited phonons and the appearance of non-oscillatory components, which occurs almost synchronously with the application of the THz field. This was confirmed in the short-time Fourier transformed spectra as shown in Fig. S11, where the excitation of most phonons was observed just after the time zero (Here we assign the modes according to the discussion in the Main Text). The AFD mode ($A_{1g}$) is however an exception which appeared at 1.5 ps and lasted much longer than the others. This is well explained by the transition pathway of $SrTiO_3$. During the application of a single-pulse THz waveform, the transient electric field on the sample first increases and then reaches the peak value of 1000 kV/cm. Therefore, the phonon modes not related to the hQPE phase can be highly excited at this stage, as is the case for most modes. The transition into the hQPE phase, however, happens somewhat later, relying not only on a sufficient transient amplitude of the electric field but also on the redistribution of atoms. This clearly associates the large driven amplitude of the $A_{1g}$ mode with the hQPE phase. It is also noted that the $A_{1g}$ mode cannot be directly driven by the THz field, and is very likely excited through a nonlinear coupling with the FES mode, which again puts the excitation of the $A_{1g}$ mode into a slower regime.

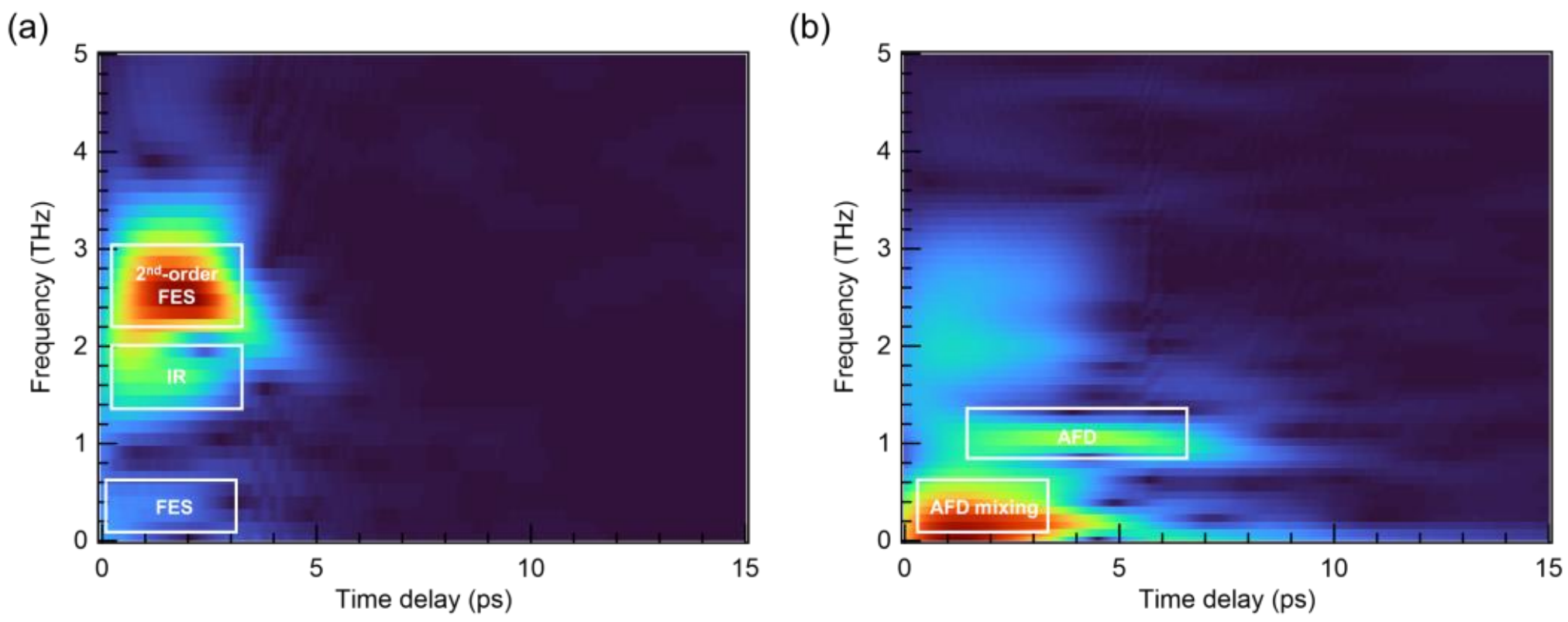

FIG. S11. Short-time Fourier transformed spectra of the experiment results. (a) The TFISH response at 1000 kV/cm. (b) The TKE response at 1000 kV/cm. The modes are assigned according to the discussion in the Main Text.